\documentstyle[aps,prl,multicol,epsfig]{revtex}
\newcommand{\bq}{\begin{equation}}
\newcommand{\eq}{\end{equation}}
\newcommand{\bqa}{\begin{eqnarray}}
\newcommand{\eqa}{\end{eqnarray}}
\newcommand{\nn}{\nonumber \\}

\begin{document}
\draft 
\title{High $T_c$ Phase Diagram based on the SU(2) Slave-Boson Approach to the t-J Hamiltonian}
\author{Sung-Sik Lee and Sung-Ho Suck Salk$^a$}
\address{Department of Physics, Pohang University of Science and Technology,\\
Pohang, Kyoungbuk, Korea 790-784\\
$^a$ Korea Institute of Advanced Studies, Seoul 130-012, Korea\\}
\date{\today}

\maketitle

\begin{abstract}
Based on an improved SU(2) slave-boson approach to the t-J Hamiltonian, we derive a phase diagram of high $T_c$ cuprates which displays both the superconducting and pseudogap phases in the plane of temperature vs. hole doping rate.
It is shown that the inclusion of phase fluctuations in the order parameters results in a closer agreement with the observed superconducting temperature(bose condensation line) of an arch shape as a function of hole doping rate.

\end{abstract}
\begin{multicols}{2}

\newpage
High $T_c$ superconductivity arises as a consequence of hole(or electron) doping in the parent cuprate oxides which are Mott insulators with antiferromagnetic long-range order.
The observed phase diagram\cite{YASUOKA}\cite{ODA} in the plane of temperature $T$ vs. hole doping rate $\delta$ shows the bose condensation(superconducting temperature) curve of an 'arch' shape rather than the often predicted linear increase, by manifesting the presence of the optimal doping rate of $\delta = 0.16$ to $0.2$.
On the other hand, the observed pseudogap temperature displays nearly a linear decrease with $\delta$.
Various U(1) slave-boson approaches to the t-J Hamiltonian were able to predict such a linear decrease in the pseudogap temperature as a function of $\delta$\cite{KOTLIAR}-\cite{GIMM}.
On the other hand, these theories failed to predict the experimentally observed bose condensation temperature $T_c$ of the arch shape as a function of $\delta$.
Instead a linear increase of $T_c$ with $\delta$ was predicted.
Further the pseudogap phase was shown to disappear when the gauge fluctuations are introduced into the U(1) slave-boson mean field theory\cite{UBBENS}.
Most recently Wen and Lee proposed an SU(2) theory to readily estimate the low energy phase fluctuations of order parameters and made a brief discussion on the possibility of holon(boson) pair condensation\cite{WEN}.
In view of failure in the correct prediction of the bose condensation temperature $T_c$ in the phase diagram with earlier theories, in the present study we examine the variation of the holon-pair condensation temperature with the hole doping rate, by treating the phase fluctuations of the order parameters in the SU(2) slave boson theory.
The present work differs from our previous U(1) slave-boson study(of the phase diagram involving the holon-pair bose condensation) and other earlier studies\cite{KOTLIAR}-\cite{UBBENS}(involving the single holon condensation) in that coupling between the holon and spinon degrees of freedom in the slave-boson representation of the Heisenberg term of the t-J Hamiltonian is no longer neglected.
We find from the treatment of the coupling that the predicted phase diagram displays the arch-shaped bose condensation curve(temperature $T_c$) as a function of hole doping rate in both treatments of the U(1) and SU(2) slave-boson approaches.
In addition, comparison between the two approaches will be made to reveal the importance of the low energy phase fluctuations of the order parameters.
It is noted that such phase fluctuations are not taken into account in the usual treatment of the U(1) slave-boson mean field theory.

We write the t-J Hamiltonian, 
\begin{eqnarray}
H & = & -t\sum_{<i,j>}(c_{i\sigma}^{\dagger}c_{j\sigma} + c.c.) + J\sum_{<i,j>}({\bf S}_{i} \cdot {\bf S}_{j} - \frac{1}{4}n_{i}n_{j}).
\label{eq:tjmodel1}
\end{eqnarray}
Here ${\bf S}_{i}$ is the electron spin operator at site $i$, ${\bf S}_{i}=\frac{1}{2}c_{i\alpha}^{\dagger} \bbox{\sigma}_{\alpha \beta}c_{i\beta}$ with $\bbox{\sigma}_{\alpha \beta}$, the Pauli spin matrix element and $n_i$, the electron number operator at site $i$, $n_i=c_{i\sigma}^{\dagger}c_{i\sigma}$.
We note that  ${\bf S}_{i} \cdot {\bf S}_{j} - \frac{1}{4}n_{i}n_{j} =  -\frac{1}{2} ( c_{i2}^{\dagger}c_{j1}^{\dagger}-c_{i1}^{\dagger}c_{j2}^{\dagger}) (c_{j1}c_{i2}-c_{j2} c_{i1})$ leads to $-\frac{1}{2} b_i b_j b_j^\dagger b_i^\dagger (f_{i \downarrow}^\dagger f_{j \uparrow}^\dagger - f_{i \uparrow}^\dagger f_{j \downarrow}^\dagger ) ( f_{j \uparrow} f_{i \downarrow} - f_{j \downarrow} f_{i \uparrow} )$ in the U(1) slave boson representation.
In earlier studies of the slave-boson theory, it is often assumed that $b_i b_j b_j^\dagger b_i^\dagger=1$.
Strictly speaking, this is precise only at half-filling(or no hole doping) and may be approximately valid near half-filling(or sufficiently low hole doping), where charge fluctuations at each site can be ignored owing to the hindrance of electron hopping from site to site.
Thus, coupling between the holon($b$) and spinon($f$) degrees of freedom has been neglected in the earlier treatments of the Heisenberg term.
Including our recent work\cite{GIMM}, the Heisenberg term was approximately treated by ignoring coupling between the spinon($f$) and holon($h$) degrees of freedom.
By allowing the coupling in the SU(2) slave-boson representation\cite{WEN}, the t-J Hamiltonian above can be written,
\bqa
H  & =  &  - \frac{t}{2} \sum_{<i,j>\sigma}  \Bigl[ (f_{\sigma i}^{\dagger}f_{\sigma j})(b_{1j}^{\dagger}b_{1i}-b_{2i}^{\dagger}b_{2j}) \nn
 && + (f_{\sigma j}^{\dagger}f_{\sigma i})(b_{1i}^{\dagger}b_{1j}-b_{2j}^{\dagger}b_{2i}) \nn
&& + (f_{2i}f_{1j}-f_{1i}f_{2j}) (b_{1j}^{\dagger}b_{2i} + b_{1i}^{\dagger}b_{2j}) \nn
&& + (f_{1j}^{\dagger}f_{2i}^{\dagger}-f_{2j}^{\dagger}f_{1i}^{\dagger}) (b_{2i}^{\dagger}b_{1j}+b_{2j}^{\dagger}b_{1i}) \Bigr] \nn
 & - & \frac{J}{2} \sum_{<i,j>} ( 1 - h_{i}^\dagger h_{i} ) ( 1 - h_{j}^\dagger h_{j} ) \times \\
 && (f_{2i}^{\dagger}f_{1j}^{\dagger}-f_{1i}^ {\dagger}f_{2j}^{\dagger})(f_{1j}f_{2i}-f_{2j} f_{1i})  \nn
&& -  \mu_0 \sum_i ( h_i^\dagger h_i - \delta )  -  \sum_i  \Bigl[
 i\lambda_{i}^{(1)} ( f_{1i}^{\dagger}f_{2i}^{\dagger} + b_{1i}^{\dagger}b_{2i}) \nn
 && + i \lambda_{i}^{(2)} ( f_{2i}f_{1i} + b_{2i}^\dagger b_{1i} ) \nn
 && + i \lambda_{i}^{(3)} ( f_{1i}^{\dagger}f_{1i} -  f_{2i} f_{2i}^{\dagger} + b_{1i}^{\dagger}b_{1i} - b_{2i}^{\dagger}b_{2i} ) \Bigr].
\label{eq:tjmodel}
\eqa
Here $f_{\alpha i}$ ( $f_{\alpha i}^{\dagger}$ ) is the spinon annihilation(creation) operator 
and $h_i \equiv \left( \begin{array}{c}  b_{1i} \\ b_{2i} \end{array} \right)$ $\left( h_i^{\dagger} = (b_{1i}^{\dagger}, b_{2i}^\dagger) \right) $, the doublet of holon annihilation(creation) operators.
$\lambda_{i}^{(1),(2),(3)}$ are the real Lagrangian multipliers to enforce the local single occupancy constraint in the SU(2) slave-boson representation\cite{WEN}.

The Heisenberg interaction term(the second term in Eq.(\ref{eq:tjmodel})) above can be decomposed into terms involving mean fields and fluctuations respectively,
\bqa 
\lefteqn{ -\frac{J}{2} ( 1 - h_i^\dagger h_i) ( 1 -  h_j^{\dagger}h_j ) (f_{2i}^{\dagger}f_{1j}^{\dagger}-f_{1i}^ {\dagger}f_{2j}^{\dagger})(f_{1j}f_{2i}-f_{2j}f_{1i})}  \nn
 & = & -\frac{J}{2} \Bigl< ( 1 - h_i^\dagger h_i) ( 1 -  h_j^{\dagger}h_j ) \Bigr> \times \\
 && (f_{2i}^{\dagger}f_{1j}^{\dagger}-f_{1i}^ {\dagger}f_{2j}^{\dagger})(f_{1j}f_{2i}-f_{2j} f_{1i}) \nn
 & & -\frac{J}{2} \Bigl< (f_{2i}^{\dagger}f_{1j}^{\dagger}-f_{1i}^ {\dagger}f_{2j}^{\dagger})(f_{1j}f_{2i}-f_{2j} f_{1i}) \Bigr> \times \\
 && ( 1 - h_i^\dagger h_i) ( 1 -  h_j^{\dagger}h_j ) \nn
 & & + \frac{J}{2} \Bigl< ( 1 - h_i^\dagger h_i) ( 1 -  h_j^{\dagger}h_j ) \Bigr> \times \\
 && \Bigl< (f_{2i}^{\dagger}f_{1j}^{\dagger}-f_{1i}^ {\dagger}f_{2j}^{\dagger})(f_{1j}f_{2i}-f_{2j} f_{1i}) \Bigr> \nn
  & & -\frac{J}{2} \Bigl( ( 1 - h_i^\dagger h_i) ( 1 -  h_j^{\dagger}h_j ) - \Bigl< ( 1 - h_i^\dagger h_i) ( 1 -  h_j^{\dagger}h_j ) \Bigr> \Bigr) \times \nn
  && \Bigl( (f_{2i}^{\dagger}f_{1j}^{\dagger}-f_{1i}^ {\dagger}f_{2j}^{\dagger} ) ( f_{1j}f_{2i}-f_{2j} f_{1i}) \nn
  && - \Bigl<(f_{2i}^{\dagger}f_{1j}^{\dagger}-f_{1i}^ {\dagger}f_{2j}^{\dagger})(f_{1j}f_{2i}-f_{2j} f_{1i})\Bigr> \Bigr). \label{eq:mf_fluc} \nn
\eqa

By introducing the Hubbard-Stratonovich fields, ${\rho}_{i}^{k}$, $\chi_{ij}$ and $\Delta_{ij}$ in association with the direct, exchange and pairing channels of the spinon, we obtain the effective Hamiltonian from Eq.(\ref{eq:tjmodel}),
\begin{eqnarray}
\lefteqn{H_{eff}  = } \nn
&& \frac{J(1-\delta)^2}{2} \sum_{<i,j>} \sum_{l=0}^{3} \Bigl( (\rho^l_{ij})^2 - \rho^l_{ij} ( f_i^{\dagger} \sigma^l f_i ) \Bigr) \nn
 & + & \frac{J(1-\delta)^2}{4} \sum_{<i,j>} \Bigl[ |\chi_{ij}|^2 - \{ f_{\sigma i}^{\dagger}f_{ \sigma j} \nn
 && + \frac{2t}{J(1-\delta)^2} (b_{1i}^{\dagger}b_{1j}-b_{2j}^{\dagger}b_{2i }) \} \chi_{ij} - c.c. \Bigr] \nn
 & + & \frac{J(1-\delta)^2}{2} \sum_{<i,j>} \Bigl[ |\Delta_{ij}|^2 - \{ (f_{2i}^{\dagger}f_{1j}^{\dagger}-f_{1i}^{\dagger}f_{2j}^{\dagger}) \nn
 && - \frac{t}{J(1-\delta)^2} (b_{1j}^{ \dagger}b_{2i} + b_{1i}^{\dagger}b_{2j}) \} \Delta_{ij} - c.c. \Bigr]  \nn
 &  - &  \frac{J}{2} \sum_{<i,j>} |\Delta^f_{ij}|^2 \Bigl[ \sum_{\alpha,\beta} b_{\alpha i}^\dagger b_{\beta j}^\dagger b_{\beta j} b_{\alpha i}  - ( h_j^{\dagger} h_j + h_{i}^\dagger h_{i}  - 2\delta ) - \delta^2 \Bigr] \nn
 & + &  \frac{t^2}{J(1-\delta)^2}  \sum_{<i,j>} \Bigl[ (b_{1i}^{\dagger}b_{1j}-b_{2j}^{\dagger}b_{2i}) (b_{1j}^{\dagger}b_{1i}-b_{2i}^{\dagger}b_{2j}) \nn
 && + \frac{1}{2} (b_{1j}^{\dagger}b_{2i}+b_{1i}^{\dagger}b_{2j}) (b_{2i}^{\dagger}b_{1j} + b_{2j}^{\dagger}b_{1i}) \Bigr]  \nn
 & + & \frac{J(1-\delta)^2}{2} \sum_{i,\sigma} (f_{\sigma i}^\dagger f_{\sigma i})
-  \mu_0 \sum_i ( h_i^\dagger h_i - \delta ) \nn
& - & \sum_i  \Bigl[
 i \lambda_{i}^{1} ( f_{1i}^{\dagger}f_{2i}^{\dagger} + b_{1i}^{\dagger}b_{2i})
+ i \lambda_{i}^{2} ( f_{2i}f_{1i} + b_{2i}^\dagger b_{1i} ) \nn
&& + i \lambda_{i}^{3} ( f_{1i}^{\dagger}f_{1i} -  f_{2i} f_{2i}^{\dagger} + b_{1i}^{\dagger}b_{1i} - b_{2i}^{\dagger}b_{2i} )], 
\label{eq:mf_hamiltonian1}
\end{eqnarray}
where $\Delta_{ij} = \Bigl< (f_{1i}f_{2j} - f_{2i}f_{1j}) - \frac{t}{J(1-\delta)^2} ( b_{2i}^\dagger b_{1j} + b_{2j}^\dagger b_{1i} ) \Bigr> = \Delta_{ij}^f - \frac{t}{J(1-\delta)} \chi_{ij;12}^b$, with $\chi_{ij;12}^b = \Bigl< b_{2i}^\dagger b_{1j} + b_{2j}^\dagger b_{1i} \Bigr>$ with $\delta$, hole doping rate.
In Eq.(\ref{eq:mf_hamiltonian1}) above we introduced $\Bigl< (f_{2i}^{\dagger}f_{1j}^{\dagger}-f_{1i}^ {\dagger}f_{2j}^{\dagger})(f_{1j}f_{2i}-f_{2j} f_{1i}) \Bigr> \approx \Bigl< (f_{2i}^{\dagger}f_{1j}^{\dagger}-f_{1i}^ {\dagger}f_{2j}^{\dagger}) \Bigr> \Bigl< (f_{1j}f_{2i}-f_{2j} f_{1i}) \Bigr> = |\Delta^f_{ij}|^2$ and $ \Bigl< ( 1 - h_i^\dagger h_i) ( 1 -  h_j^{\dagger}h_j ) \Bigr> \approx \Bigl< ( 1 - h_i^\dagger h_i)\Bigr> \Bigl< ( 1 -  h_j^{\dagger}h_j ) \Bigr> = ( 1-\delta)^2$ and neglected the last term in Eq.(\ref{eq:mf_fluc}) above.

The four boson term in the fourth term of Eq.(\ref{eq:mf_hamiltonian1}) allows holon pairing and a scalar boson field, $\Delta_{ij ;\alpha \beta}^b$ is introduced for the holon pairing between the nearest neighbor $b_{\alpha}-$ and $b_{\beta}-$single bosons with the boson index, $\alpha, \beta =$ $1$ or $2$\cite{WEN}.
Using the saddle point approximation, we obtain from Eq.(\ref{eq:mf_hamiltonian1}) the mean field Hamiltonian,
\begin{eqnarray}
\lefteqn{ H^{MF}= \frac{J(1-\delta)^2}{2} \sum_{<i,j>} \Bigl[ |\Delta_{ij}^{f}|^{2} + \frac{1}{2} |\chi_{ij}|^{2} + \frac{1}{4} \Bigr] } \nn
&& + \frac{J}{2} \sum_{<i,j>} |\Delta^f_{ij}|^2 \Bigl[ \sum_{\alpha,\beta} |\Delta_{ij; \alpha \beta}^{b}|^{2} + \delta^2  \Bigr]   \nn
& & -\frac{J(1-\delta)^2}{2} \sum_{<i,j>} \Bigl[ \Delta_{ij}^{f*} (f_{1j}f_{2i}-f_{2j}f_{1i}) + c.c. \Bigr] \nn
&& -\frac{J(1-\delta)^2}{4} \sum_{<i,j>} \Bigl[ \chi_{ij} (f_{\sigma i}^{\dagger}f_{\sigma j}) + c.c. \Bigr] + \nonumber \\
& & -\frac{t}{2} \sum_{<i,j>} \Bigl[ \chi_{ij}(b_{1i}^{\dagger}b_{1j} - b_{2j}^{\dagger}b_{2i}) -\Delta^f_{ij} (b_{1j}^{\dagger}b_{2i} + b_{1i}^{\dagger}b_{2j})\Bigr] - c.c. \nn
&& -\sum_{<i,j>,\alpha,\beta} \frac{J}{2}|\Delta^f_{ij}|^2 \Bigl[ \Delta_{ij;\alpha \beta }^{b*} (b_{\alpha i}b_{\beta j}) + c.c. \Bigr]  \nn
&& -\sum_{i} \Bigl[ \mu_{i} ( h_{i}^{\dagger}h_{i} - \delta )  
+ i\lambda_{i}^{1} ( f_{1i}^{\dagger}f_{2i}^{\dagger} + b_{1i}^{\dagger}b_{2i}^{\dagger}) \nn
&& + i \lambda_{i}^{2} ( f_{2i}f_{1i} + b_{2i}b_{1i} )
+ i \lambda_{i}^{3} ( f_{1i}^{\dagger}f_{1i} -  f_{2i} f_{2i}^{\dagger} + b_{1i}^{\dagger}b_{1i} + b_{2i}^{\dagger}b_{2i} ) \Bigr] \nn
&& - \frac{t}{2} \sum_{<i,j>}  \left( \Delta^{f}_{ij} - (f_{1j}f_{2i}- f_{2j}f_{1i}) \right) \chi_{ij;12}^{b*} - c.c. \nn
&& + \frac{t^2}{2J(1-\delta)^2} \sum_{<i,j>}  \left| \chi_{ij;12}^b - (b_{2i}^\dagger b_{1j} + b_{2j}^\dagger b_{1i} ) \right|^2  \nn
&& + \frac{t^2}{J(1-\delta)^2}\sum_{<i,j>}  (b_{1i}^{\dagger}b_{1j} - b_{2j}^{\dagger}b_{2i}) ( b_{1j}^{\dagger}b_{1i} - b_{2i}^{\dagger}b_{2j}),
\label{eq:mf_hamiltonian2}
\end{eqnarray}
where $\chi_{ij}= < f_{\sigma j}^{\dagger}f_{\sigma i} + \frac{2t}{J(1-\delta)^2} (b_{1j}^{\dagger}b_{1i} - b_{2i}^\dagger b_{2j} )>$, $\Delta_{ij}^{f}=< f_{1j}f_{2i}-f_{2j}f_{1i} >$, $\Delta_{ij;\alpha\beta}^{b} = <b_{i\alpha}b_{\beta j}>$ and $\mu_i = \mu_0 - \frac{J}{2}\sum_{j=i\pm \hat x, i \pm \hat y} |\Delta^f_{ij}|^2$.
The Hubbard Stratonovich field $\rho_{i}^{k=1,2,3}=<\frac{1}{2}f_{i}^\dagger \sigma^k f_i>$ for direct channel is taken to be $0$\cite{UBBENS} and $\rho_i^{k=0}=\frac{1}{2}$.
Owing to the energy cost the exchange interaction terms(the last two positive energy terms in Eq.(\ref{eq:mf_hamiltonian2})) is usually ignored\cite{UBBENS}-\cite{WEN}.

We now introduce the uniform hopping order parameter, $\chi_{ij}=\chi$, the d-wave spinon pairing order parameter, $ \Delta_{ij}^{f}=\pm \Delta_f$ with the sign $+(-)$ for the nearest neighbor link parallel to $\hat x$ ($\hat y$) and the s-wave holon pairing order parameter, $\Delta_{ij;\alpha \beta}^{b}=\Delta^b_{ \alpha \beta}$ with the boson indices $\alpha$ and $\beta$.
For the case of $\Delta^b_{\alpha \beta}=0$, $\lambda^{(k)}=0$ and $\Delta^f \leq \chi$, the $b_1$-bosons are populated at and near $k=(0,0)$ in the momentum space and the $b_2$-bosons, at and near $k=(\pi,\pi)$\cite{WEN}.
Pairing of two different($\alpha \neq \beta$) bosons(holons) gives rise to the non-zero center of mass momentum.
On the other hand, the center of mass momentum is zero only for pairing between identical($\alpha = \beta$) bosons.
Thus writing $\Delta^b_{ \alpha \beta} = \Delta_b ( \delta_{\alpha,1}\delta_{\beta,1} - \delta_{\alpha,2} \delta_{\beta,2} )$\cite{WEN} for pairing between the identical holons and allowing the uniform chemical potential, $\mu_{i}=\mu$, the mean field Hamiltonian from Eq.(\ref{eq:mf_hamiltonian2}) is derived to be,
\begin{eqnarray}
H^{MF} & = & N J (1-\delta)^2 \Bigl( \frac{1}{2}\chi^{2} + \Delta_f^{2} + \frac{1}{4} \Bigr) + NJ\Delta_f^2 ( 2\Delta_b^2 + \delta^2 ) \nn
& + & \sum_{k} E_{k}^{f} (\alpha_{k1}^{\dagger}\alpha_{k1} - \alpha_{k2}\alpha_{k2}^{\dagger}) \nn
& + &  \sum_{k,s=1,2} \Bigl[ E_{ks}^{b} \beta_{ks}^{\dagger}  \beta_{ks}  +  \frac{1}{2}( E_{ks}^b + \mu ) \Bigr] + \mu N \delta.
\label{eq:diagonalized_hamiltonian}
\end{eqnarray}
Here $E_{k}^{f}$ and $E_{ks}^{b}$ are the quasiparticle energies of spinon and holon respectively.
$\alpha_{ks}( \alpha_{ks}^{\dagger})$ and $\beta_{ks}(\beta_{ks}^{\dagger})$ are the annihilation(creation) operators of the spinon quasiparticles and the holon quasiparticles respectively.

From the diagonalized Hamiltonian Eq.(\ref{eq:diagonalized_hamiltonian}), we readily obtain the total free energy, 
\begin{eqnarray}
F & = &  NJ(1-\delta)^2 \Bigl( \frac{1}{4} + \Delta_f^{2} + \frac{1}{2}\chi^{2} \Bigr)  \nn
 && - 2k_{B}T \sum_{k} ln [ \cosh (\beta E_{k}^{f}/2) ] \nonumber \\
&& + NJ\Delta_f^2 ( \Delta_b^{2} + \delta^2 ) + k_{B}T \sum_{k,s} ln [1 - e^{-\beta E_{ks}^{b}}] \nn
&& + \sum_{k,s} \frac{ E_{ks}^{b} + \mu }{2}  + \mu N \delta.
\label{eq:free_energy}
\end{eqnarray}
The chemical potential is determined from the number constraint of doped holes,
\bqa
\lefteqn{- \frac{\partial F}{\partial \mu}  =   \sum_k \Bigl[  \frac{1}{e^{\beta E_{k1}^b} -1}\frac{-\epsilon_k^b - \mu}{E_{k1}^b} + \frac{1}{2}(\frac{-\epsilon_k^b-\mu}{E_{k1}^b} - 1 )   } \nn
 && +   \frac{1}{e^{\beta E_{k2}^b} -1}\frac{\epsilon_k^b - \mu}{E_{k2}^b} + \frac{1}{2}( \frac{ \epsilon_k^b-\mu }{ E_{k2}^b }  - 1 )    \Bigr] - N\delta  =  0, \label{mu_eq} 
\eqa
and the Lagrangian multipliers are determined by the following three constraints imposed by the SU(2) slave-boson theory, 
\bqa
\lefteqn{ \frac{\partial F}{ \partial \lambda^{(k)} }  =  
-\sum_k \tanh \frac{ \beta E_k^f }{2} \frac{ \partial E_k^f }{ \partial \lambda^{(k)} } } \nn
&& + \sum_{k,s} \frac{ e^{\beta E_{ks}^b} + 1 }{ 2(e^{\beta E_{ks}^b}-1) } \frac{ \partial E_{ks}^b }{ \partial \lambda^{(k)} } = 0 , \mbox{ $k=1,2,3$} \label{constraint_eq}.
\eqa
It can be readily proven from Eq.(\ref{constraint_eq}) above that $\lambda^{(k)}=0$ satisfies the three constraints above. 

By minimizing the free energy, the order parameters $\chi$, $\Delta_f$ and $\Delta_b$ are numerically determined as a function of temperature and doping rate.
In Fig.1 the mean field results of the U(1) slave-boson theory(dotted lines) are displayed for $J=0.2$ $t$, $J=0.3$ $t$ and $J=0.4$ $t$ for comparison with the predicted phase diagrams(solid lines).
The predicted pseudogap(spin gap) temperature, $T^f_{SU(2)}$ is consistently higher than $T^f_{U(1)}$, the U(1) value.
We note from the fourth term in Eq.(\ref{eq:mf_hamiltonian1}) that the holon pairing channel depends on the spinon pairing order parameter $\Delta_f$.
Accordingly the predicted holon pair condensation temperature(superconducting transition temperature) $T^b_{SU(2)}$ depends on the spin gap(pseudogap) temperature $T^*$; $T^b_{SU(2)}$ in the overdoped region decreases with $T^*$.
$T^b_{SU(2)}$ at optimal doping is predicted to be lower than the value of $T^b_{U(1)}$ predicted by the U(1) theory.
The predicted optimal doping rate is shifted to a larger value, showing closer agreement with observation\cite{YASUOKA}\cite{ODA} than the U(1) mean field treatment.
Such discrepancies are attributed to the phase fluctuations of order parameters, which were not treated in the U(1) mean field theory.

In summary, based on the SU(2) slave-boson symmetry conserving t-J Hamiltonian we derived a phase diagram of high $T_c$ cuprates which displays the bose condensation temperature of an arch shape as a function of hole doping rate. 
Unlike other previous studies which predicted a linear increase with the hole doping rate, this result is consistent with observation.
We showed that the low energy fluctuations cause a shift of the optimal doping rate to a larger value and a suppression of the holon pair bose condensation temperature, allowing closer agreement with observation compared to the U(1) case.

One(SHSS) of us acknowledges the generous supports of Korea Ministry of Education(BSRI-98) and the Center for Molecular Science at Korea Advanced Institute of Science and Technology.
We thank Tae-Hyoung Gimm for helpful discussions.

\begin{minipage}[c]{9cm}
\begin{figure}
\vspace{0cm}
\epsfig{file=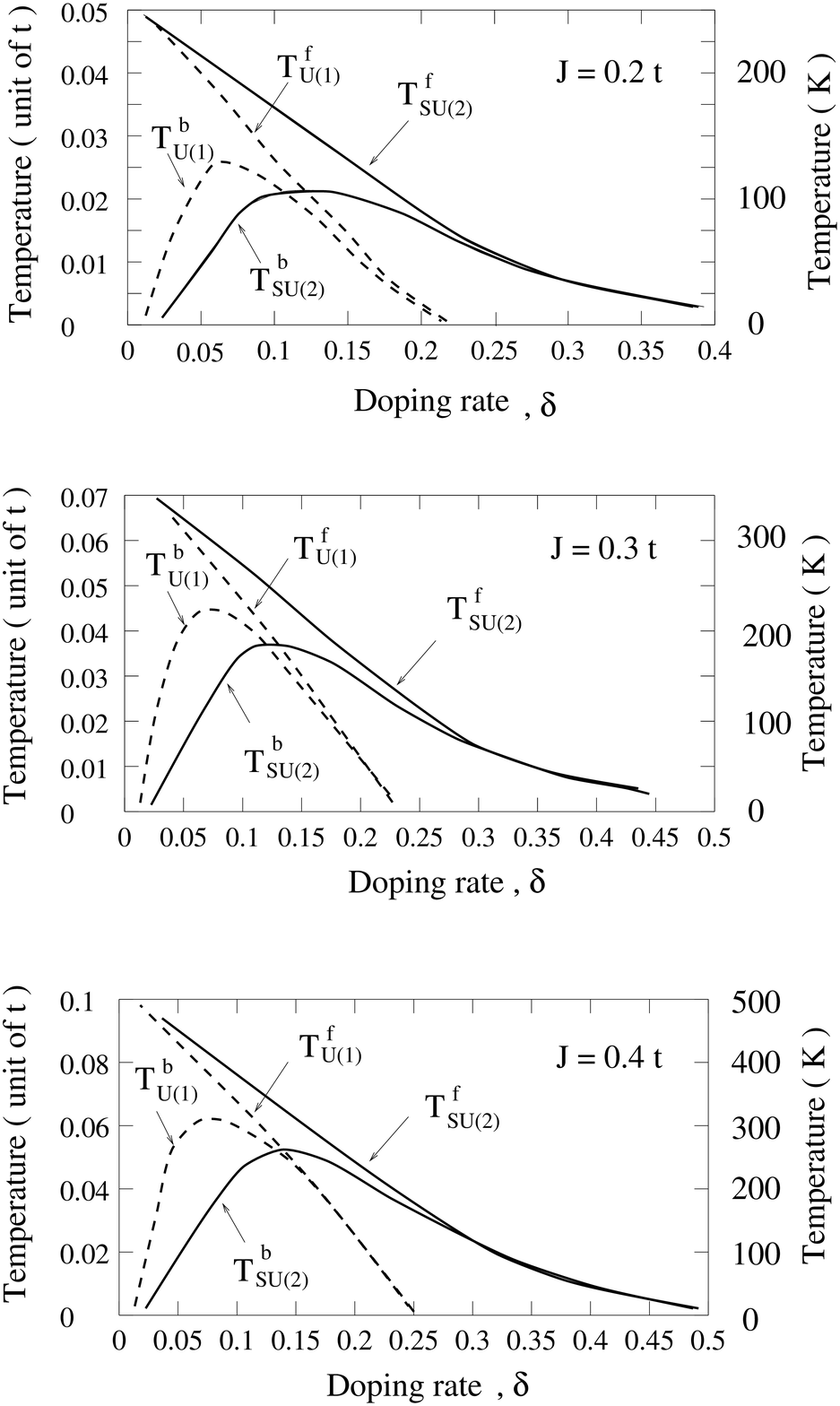,angle=0, height=10cm, width=7cm}
\label{fig:1}
\caption{
Computed phase diagrams with $J=0.2t$, $J=0.3t$ and $J=0.4t$.
$T^f_{SU(2)}$($T^f_{U(1)}$) denotes the pseudogap temperature  and $T^b_{SU(2)}$($T^b_{U(1)}$), the holon pair bose condensation temperature predicted from the SU(2)(solid lines) and (U(1))(dotted lines) slave-boson theories respectively.
 The scale of temperature in the figure is based on $t=0.44eV$[5].
}
\end{figure}
 \end{minipage}


\end{multicols}
\end{document}